\documentclass[floatfix,aps,prl,groupedaddress, twocolumn]{revtex4-2}
\usepackage{braket}
\usepackage{amsmath}
\usepackage{svg}
\usepackage[utf8]{inputenc}
\usepackage[T1]{fontenc}
\usepackage{lineno}
\usepackage{comment}
\usepackage{graphicx}
\usepackage[normalem]{ulem}
\usepackage[colorlinks,breaklinks]{hyperref}
\usepackage{mathtools}

\definecolor{darkred}{rgb}{0.5,0,0}
\definecolor{darkgreen}{rgb}{0,0.5,0}
\definecolor{darkblue}{rgb}{0,0,0.5}
\hypersetup{linkcolor=darkred,citecolor=darkgreen,urlcolor=darkblue}

\makeatletter
\renewcommand{\p@subsection}{}
\renewcommand{\p@subsubsection}{}
\makeatother

\usepackage[toc,page]{appendix}
\usepackage{tikz}
\usetikzlibrary{math}
\usetikzlibrary{decorations.pathreplacing,calligraphy}

\begin{document}

\title{
Orthogonalised Self-Guided Quantum Tomography: Insights from Single-Pixel Imaging}

\author{Kiki Dekkers$^{1*}$, Alice Ruget$^{1*}$, Fazilah Nothlawala$^{2}$, Sabrina Henry$^{1}$, Stirling Scholes$^{1}$, Miles Padgett$^{3}$, Andrew Forbes$^{2}$, Isaac Nape$^{2}$, and Jonathan Leach$^{1,\, \dagger}$}

\affiliation{$^{1}$School of Engineering and Physical Sciences, Heriot-Watt University, Edinburgh, EH14 4AS, UK}
\affiliation{$^{2}$School of Physics, University of the Witwatersrand, Private Bag 3, Wits 2050, South Africa}
\affiliation{$^{3}$University of Glasgow, Glasgow, G12 8QQ, UK}

\thanks{These authors contributed equally}
\email{$^{\dagger}$ j.leach@hw.ac.uk}
















\begin{abstract}

\noindent    We introduce the concept of self-guided imaging (SGI) as a linear analogue of self-guided quantum tomography (SGQT). We show that SGI is mathematically equivalent to single-pixel imaging (SPI).  Taking inspiration from orthogonalised ghost imaging, a recent advance in SPI, we introduce orthogonalised SGQT. This requires no additional experimental overhead and leads to faster and more accurate final convergence, as we demonstrate numerically (fidelity $95.2\% \rightarrow 99.17\%$) and experimentally (fidelity $92.1\% \rightarrow 95.3\%$). This work suggests that further routines from SPI and SGQT can be interchanged to optimise measurements and convergence.


\end{abstract}

\maketitle

\emph{Introduction}:-- The efficient and optimal reconstruction of systems, such as images or quantum states, remains a central challenge in modern physics. These problems can be solved via appropriate algorithms such as single-pixel imaging (SPI) for the former and quantum tomography for the latter. However, due to the scaling, computational cost, and the need to solve efficiently and accurately, developing new algorithms remains an important field of research, recently reviewed in the context of methods for learning quantum systems \cite{Gebhart} and SPI \cite{Gibson:20}.  


 \begin{figure}
    \centering
    \includegraphics[width=.9\linewidth]{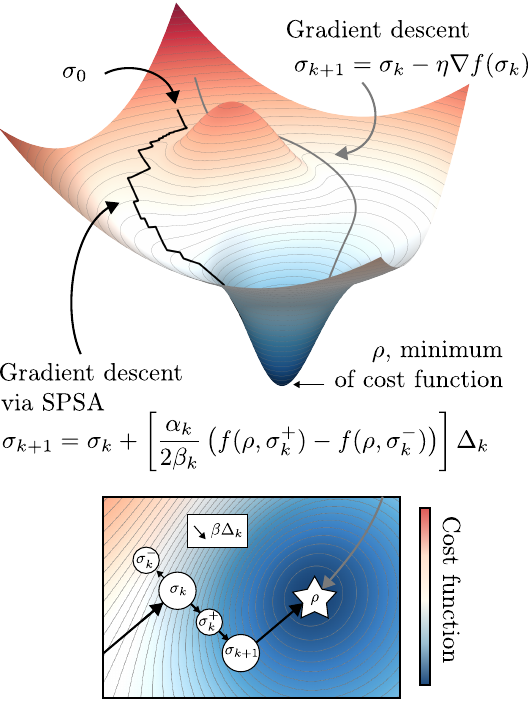}
    \caption{Illustration of gradient descent via conventional (grey line) and via the simultaneous perturbation stochastic approximation (black line). Both SGQT and versions of SPI are examples of gradient descent via the SPSA method.} 
    \label{fig:SQGT_concept}
\end{figure}

Quantum tomography is the process of establishing an unknown quantum state $\rho$ from an informationally complete set of known measurements and the corresponding outcomes \cite{MAURODARIANO2003205, RN1363, PhysRevA.64.052312, RN1365}.  An attractive alternative was developed by Ferrie, so-called self-guided quantum tomography (SGQT) \cite{RN1358}. In SGQT the current estimate guides itself, through iterative feedback directly from measurements, to the true state without explicit reconstruction. It is therefore not necessary to collect a large set of data (scaling as $2d^{2n} - 1$ for $n$ qu$d$its \cite{thew_qudit_2002}) and then solve the computationally intensive inverse problem as is the case for quantum tomography. SGQT has been implemented experimentally, for example, in the context of polarization photonic qubits \cite{PhysRevLett.117.040402} and high-dimensional quantum states \cite{Rambach_tomography_2021, serino_self-guided_2025}, and it has been extended for quantum process tomography \cite{RN1361}.

In a completely separate field to quantum tomography, single-pixel imaging (SPI) \cite{Gibson:20, 4472247} seeks to establish an unknown image from the outcomes of a set of known measurements of an object. Such systems can provide intensity and depth imaging at non-visible wavelengths where arrayed sensors do not exist.     
Although SGQT and SPI differ in spirit in many ways, in this work we show that the SPI and SGQT algorithms are equivalent in the case of a linear distance measure. We show this numerically in the context of real-valued images, introducing the concept of self-guided imaging (SGI). 
Noting the similarities between SGQT and SPI, we then draw inspiration from orthogonalised ghost imaging (OGI) \cite{kallepalli_challenging_2022}, which has recently seen great success in the context of SPI where orthogonality of measurement masks is not possible. Their work incorporates the Kaczmarz algorithm \cite{POPA2004579, strohmer2009randomized, pires2016projections, sun2020randomized} for solving systems of linear equations. We incorporate such computational addition to SGQT, introducing orthogonalised self-guided quantum tomography (OSGQT). We now demonstrate the benefits of OSGQT over SGQT.  We show this for the case of heralded single-photon states encoded in high dimensions and anticipate that the gains will be similar in other quantum systems. We show numerically and experimentally that OSGQT allows for faster convergence with higher final fidelities without any additional experimental overhead.  

\emph{Standard self-guided quantum tomography (SGQT)}:-- A brief summary of the SGQT algorithm follows.  We start with an initial guess $\ket{\sigma}$ of the unknown quantum state $\rho$.  We then make two updates to this guess labeled $\ket{\sigma^{+}}$ and $\ket{\sigma^{-}}$, where $\ket{\sigma^{\pm}}= \ket{\sigma \pm \beta \Delta}$. Here $\beta$ is a constant and $\Delta$ is a vector where each element is randomly assigned a value $\{ 1, -1, i, -i\}$.  We then make measurements $f(\rho, \sigma^{\pm})$ that determine how close these two states are. In the case of quantum measurements the distance measure can be given by the fidelity (or infidelity) $f(\rho, \sigma^{\pm}) = F(\rho, \sigma^{\pm})$ $(\textrm{or}~1-F(\rho, \sigma^{\pm}))$. 
For pure states, i.e.~$\rho = \ket{\psi}\bra{\psi}$, this simplifies to $f(\psi, \sigma^{\pm}) = F(\psi, \sigma^{\pm}) = |\langle \psi | \sigma^{\pm}\rangle|^2 $. In this work, we only consider pure states and will therefore refer to the unknown state as $\ket{\psi}$. 

The core of SGQT is that after the initial guess, the $k+1$ iteration of $\ket{\sigma}$ is given by 
\begin{align}\label{SGQT}
\ket{\sigma_{k+1}} & = \ket{\sigma_{k}+ \left[  \alpha_k \left( \frac{f(\psi, \sigma_k^+)-f(\psi, \sigma_k^-)}{2 \beta_k} \right) \right] \Delta_k}.
\end{align}
Here, $\alpha_k$ and $\beta_k$ are functions that control the convergence of the algorithm.  We can understand each iteration of SGQT intuitively by considering that the quantity in the parenthesis ($\cdot$) is the gradient $g_k$ along direction $\Delta_k$ in a cost function landscape where $f$ is the distance measure, see Fig.~\ref{fig:SQGT_concept}. 
Walking along (or opposite when $g_k$ is negative) this $\Delta_k$ direction with stepsize $\alpha_k g_k$ takes us closer to the desired solution. Note that the gradient is estimated using the simultaneous perturbation stochastic approximation (SPSA) algorithm \cite{RN1367}, requiring only two evaluations of the objective function per iteration. Additionally, the mathematical analogy of ghost-imaging and gradient descent was reported recently  \cite{Yu_gradient-descent_2021}.

\emph{Standard single-pixel imaging (SPI)}:--
SPI is the process of establishing an unknown object $\mathcal{O}$ by measuring a series of overlaps $f$ of an unknown object with known intensity or illuminating pattern $\Delta$, having used the notation from SGQT. An estimate $\sigma$ of the object can be calculated iteratively via
\begin{align}
\sigma_{k+1}  & =  \sigma_k + f(\mathcal{O}, \Delta_k) \Delta_k \nonumber \\ 
& = \sigma_k + \langle \mathcal{O}| \Delta_k \rangle \Delta_k,
\label{SGI}
\end{align}
The normalisation of the solution can be achieved at the final iteration, i.e., $\sigma = \sigma_{N}/N$.
Note that $\mathcal{O}$ and $\sigma$ represent real-, positive-valued images (pixels $\in \{0,1\}$), rather than complex-valued quantum states in SGQT. The distance measure $f(\mathcal{O}, \Delta_k) = \langle \mathcal{O} | \Delta_k\rangle$ is linear, in contrast to SGQT where $f$ corresponds to a quantum measurement that is inherently nonlinear.

\

{\emph{Self-guided imaging (SGI)}:-- }
We now consider the standard SGQT algorithm but apply it to the same problem as SPI, i.e. rather than estimating quantum state $\ket{\psi}$ we try to find real-valued image $\mathcal{O}$, henceforth referred to as self-guided imaging (SGI). That is to say, we replace the square of the sum of the complex overlap from quantum measurements, i.e.~$|\langle \psi | \sigma\rangle|^2 $, with the sum of the linear overlap associated with intensity images, i.e.~$\langle\mathcal{O} | \sigma\rangle$. 

As the distance measure is now linear, we can apply the substitution $
f(\mathcal{O}, \sigma \pm \beta \Delta) \rightarrow f(\mathcal{O}, \sigma) \pm \beta f(\mathcal{O}, \Delta)$ to Eq.~\ref{SGQT}.  We therefore find
\begin{align}\label{simplification}
& \frac{f(\mathcal{O}, \sigma^+)-f(\mathcal{O}, \sigma^-)}{2 \beta} \nonumber \\
 = & \frac{f(\mathcal{O},\sigma+\beta \Delta)-f(\mathcal{O},\sigma-\beta\Delta)}{2 \beta} \nonumber \\
\rightarrow & \frac{f(\mathcal{O},\sigma)+\beta f(\mathcal{O},\Delta )-f(\mathcal{O},\sigma)+\beta f(\mathcal{O},\Delta)}{2 \beta} \nonumber \\
\rightarrow & f(\mathcal{O},\Delta )  = \langle \mathcal{O} | \Delta \rangle.
\end{align}
Which means that we recover the standard approach to SPI other than the scaling factor $\alpha_k$:
\begin{align}\label{SPI_alpha_k}
\sigma_{k+1} & = \sigma_{k} + \alpha_k \langle \mathcal{O} | \Delta_k \rangle \Delta_k. 
\end{align} 
We see here that self-guided tomography simplifies to the case of single-pixel imaging in the case of linear distance measure applied to real-valued images, see the supplementary info for detailed figure of this concept. Note also that the distance measure in single-pixel imaging can be made nonlinear by the addition of regularisation (e.g., total variance or total curvature).  In this case, the simplification (Eq.~\ref{simplification}) would not apply but Eq.~\ref{SGQT} remains a valid solution. 

\begin{figure}[t]
    \centering
    \includegraphics{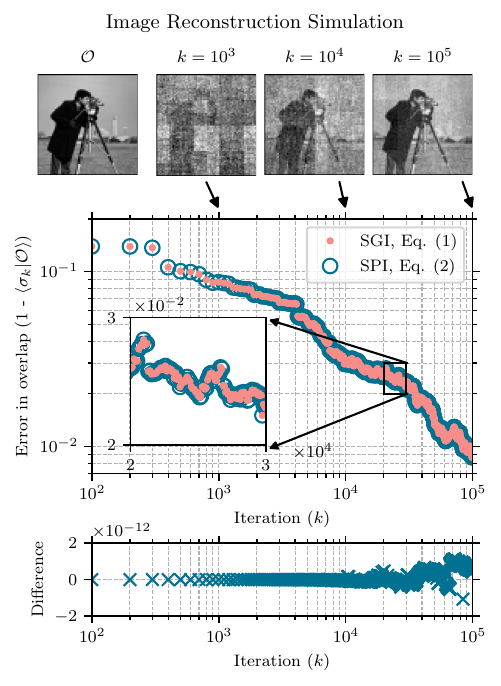}
    \caption{Comparison of error for simulation of single-pixel imaging vs.~self-guided imaging, i.e. self-guided tomography with linear distance measure, for Hadamard masks in the case of medium noise ($\gamma = 0.25$). Only every 100$^{\rm th}$ data point is shown for clarity, and the results are for a 64$\times$64 pixel image as shown at the top.}
    \label{fig:Hadamard_noise}
\end{figure}

\emph{Numerical results for SPI vs SGI}:-- Fig.~\ref{fig:Hadamard_noise} compares the output of single-pixel imaging to that of self-guided imaging for Hadamard masks applied to an image of 64$\times$64 pixels. 
Each image is normalised such that $\langle \mathcal{O} |\mathcal{O} \rangle = 1$, and each mask $\Delta_k$ is an array of size 64$\times$64, where each element  $\in \{1, -1\}$. The noise was added to the overlap term such that $\langle \mathcal{O}|\Delta_k \rangle \rightarrow \langle \mathcal{O}|\Delta_k \rangle + n_k$, where $n_k$ is a noise term drawn from a Gaussian random variable with mean 0 and width $\gamma$. Fig.~\ref{fig:Hadamard_noise} evaluates the error in the overlap of the true image and the guess image $(1-\langle \sigma_k|\mathcal{O}\rangle)$ as a function of iteration $k$ with $\gamma=0.25$. We see that at every iteration the single-pixel and self-guiding approaches are equivalent, with the difference between the two less than $\pm 2\times10^{-12}$.

\emph{Orthogonalised Ghost Imaging (OGI)}:--
It is known that in SPI, the fastest convergence is ensured by using orthogonal masks, such as Hadamard patterns. Orthogonal masks form a complete and independent basis and thus allow for non-redundant sampling as each SPI iteration contributes a maximum amount of unique information towards reconstructing $\mathcal{O}$.
In certain cases, it may not be possible to choose orthogonal masks $\{\Delta_k\}$. Ghost imaging \cite{pittman_optical_1995, bennink_two_photon_2002, Erkmen:10} using correlated states of light is one example of this. It was noted that when this occurs, the Kaczmarz method \cite{POPA2004579, strohmer2009randomized, pires2016projections, sun2020randomized} for solving a system of linear equations can be adopted for the case of ghost imaging or SPI \cite{kallepalli_challenging_2022}. In that work, the authors introduce the concept of orthogonalised ghost imaging (OGI), where the standard SPI approach in Eq.~\ref{SGI} is modified to include a correction term $f(\sigma_k, \Delta_k) = \langle \sigma_k | \Delta_k \rangle$. This overlap quantifies how much new information the current mask $\Delta_k$ contributes relative to the present estimate $\sigma_k$. By including this correction term, updates to $\sigma_k$ are suppressed when the current mask offers little new information, reducing the risk of overshooting within the cost function landscape and leading to faster convergence. The estimate image is then found via
%
\begin{align}
\sigma_{k+1}  = \sigma_k + \left(\langle \mathcal{O} | \Delta_k \rangle - \langle \sigma_k | \Delta_k \rangle \right) \Delta_k
\label{SGI}
\end{align}
%
We see initially in the early stages of any reconstruction (when $k$ is small) that this term is dominated by $\langle \mathcal{O} | \Delta_k \rangle$ as $\langle \sigma_k | \Delta_k \rangle \approx 0$. But as the algorithm progresses and $\sigma_k \rightarrow \mathcal{O}$, we see that the difference between the $\mathcal{O}$ and $\sigma_k$ contributions will tend to zero. \\

\begin{figure*}
    \centering
    \includegraphics{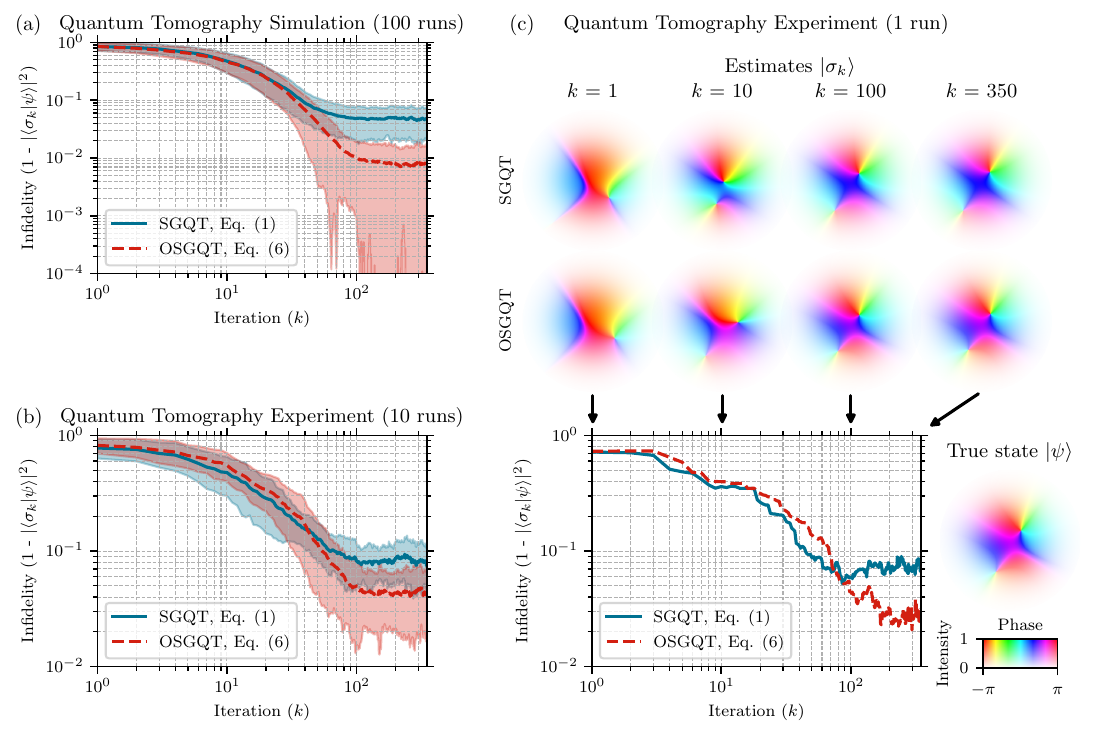}
    \caption{Comparison of self-guided quantum tomography performance for the SGQT (solid blue) and OSGQT (dashed red). Unknown quantum states were $d=5$ OAM phase qudits, i.e. $\ket{\psi} = \sum_{\ell=-2}^2c_\ell\ket{\ell}.$ (a) ((b)) Average infidelity of 100 (10) runs for no-noise simulation (experiment). Shaded regions indicate the standard deviations. Numerical and experimental results both demonstrate OSGQT reaches sub 0.1 infidelities faster on average and   achieves infidelities below the convergence saturation limit of SGQT.
    (c) Single experimental run. The phase of the current best estimates $\ket{\sigma_k}$ walk towards the phase of true state $\ket{\psi}$ as the algorithms progress, with OSGQT being the most successful. The intensity envelope represents the intensity of the modes at the SLM that is coupled into the fibres, as measured by a phase-step scan, see supplementary info.}
    \label{fig: Main_results_SGQT}
\end{figure*}

\emph{Orthogonalised Self-Guided Quantum Tomography (OSGQT)}:-- Integral to SGQT is that the algorithm does not require a fixed set of measurements but that the measurements are based on the current estimate $\ket{\sigma_k}$ and a vector $\Delta_k$. As elements in $\Delta_k$ are randomly chosen from $\{1, -1, i, -i\}$, the full set of measurements $\{\ket{\sigma^+}\}$, $\{\ket{\sigma^-}\}$ are not mutually orthogonal. We have previously shown the equivalence of SPI and SGI and noted that OGI is known to improve convergence in SPI with non-orthogonal masks. It follows that SGQT can also leverage the same computational correction as OGI to gain faster convergence by updating its estimate according to
\begin{align}\label{SGQT OTG}
\ket{\sigma_{k+1}} = |\sigma_{k}+ \Bigg[  \frac{\alpha_k}{2 \beta_k} \Bigg(&\overbrace{f(\psi, \sigma_k^+)-f(\psi, \sigma_k^-)}^\text{standard SGQT measurements} \nonumber \\
- &\underbrace{f(\sigma_k, \sigma_k^+)-f(\sigma_k, \sigma_k^-)}_\text{computational correction}\Bigg) \Bigg] \Delta_k\rangle.
\end{align}
We call this new approach orthogonalised self-guided quantum tomography (OSGQT). It carries very little additional overhead in comparison to the original version as $f(\psi, \sigma_k^{\pm})$ is experimentally measured as before, and $f(\sigma, \sigma_k^{\pm})$ is evaluated numerically on any guess solution. Note considering we are now doing quantum measurements, the cost function $f$ represents the absolute square of the complex overlap and $\ket{\psi}$, $\ket{\sigma_k}$, and $\Delta_k$ are complex-valued. 

We now show the benefit of OSGQT for the case of a heralded single-photon state in high dimensions. In the experiment, the fidelity between $\ket{\psi}$ and $\ket{\sigma^\pm}$ is estimated by $f^\pm~\approx~\frac{2 N^\pm}{N^+ + N^-}$, where $N^\pm$ are the number of detection events for $\ket{\sigma^\pm}$. 
We compare SGQT and OSGQT in simulation and in experiment for the $d=5$ single-photon quantum state $\ket{\psi}$ = $\sum_{l=-2}^2c_\ell\ket{\ell}$, where $\ket{\ell}$ represents modes with a helical phase given by $e^{i\ell\phi}$ that carry $\ell$ units of orbital angular momentum~(OAM)~ \cite{yao2011orbital, allen2016optical}. 

We used spatial light modulators (SLM) to encode $\ket{\psi}$ and the complex conjugate of $\ket{\sigma_k^\pm}$ onto the phases of two photons that are entangled in OAM. The two-photon state is generated via SPDC \cite{mair2001entanglement} in a type-I ppKTP pumped by a CW laser. The coincidence count rate, recorded by a HydraHarp after fibre-coupling, is proportional the fidelity between $\ket{\psi}$ and $\ket{\sigma_k^\pm}$ and thus can be used as feedback in our cost function. For a schematic and more details on the set up see supplementary info.
We evaluate the fidelity between the current best estimate $\ket{\sigma_k}$ and $\ket{\psi}$ numerically based on their phase overlap. Here, the intensity profile is set as the product of the SLM, mode and collection mode intensity profiles. We adopt a numerical metric rather than count rates because the algorithms were able exceed the coincidence rate threshold corresponding to for when $\ket{\sigma} = \ket{\psi}$, which returns unphysical fidelity values. However, this metric neglects prepare-and-measure and alignment imperfections and thus the reported fidelities are lower than those obtained from coincidence-based estimates. For more details on the  numerical fidelity metric and coincidence rates, see supplementary info.\\
For a fair comparison of SGQT and OSGQT, we use the same constant scaling values, $\alpha = 0.05$ and $\beta = 0.2$, chosen so that both algorithms perform near-optimal experimentally (see supplementary info). We run both algorithms for the same sets of random quantum states $\{\ket{\psi}\}$ and random starting estimates $\{\ket{\sigma_0}\}$. 
Fig.~\ref{fig: Main_results_SGQT} shows the average infidelity $1 - |\braket{\sigma_k|\psi}|^2$ for SGQT and OSGQT. We observe, numerically (Fig.~\ref{fig: Main_results_SGQT}a)  and experimentally (Fig.~\ref{fig: Main_results_SGQT}b) that OSGQT reaches the fidelity that SGQT saturates at faster. More importantly, OSGQT is able to reach fidelities that are higher than is achievable by SGQT (quoting the mean and standard error: 99.17$\pm$0.09$\%$ vs 95.2$\pm$0.2$\%$ for no-noise simulation; 95.3$\pm$0.8$\%$ vs 92.1$\pm$1.1$\%$ for experiment). The saturated trends for larger $k$ suggest that similar high fidelities are not feasible for SGQT for the same parameters (e.g. $\alpha$, $\beta$, $\ket{\psi}$, experimental parameters). 
Fig.  \ref{fig: Main_results_SGQT}c illustrates how the phase of current best estimates $\ket{\sigma_k}$ walks towards the phase of true state  $\ket{\psi}$ for a single run, where the phase of final estimate $\ket{\sigma_{k=350}}$ for OSGQT most closely resembles the true state. \\
As the performance of both algorithms is sensitive to $\alpha$ and $\beta$, we ran experiments for a range of $\alpha$ and $\beta$ values, and confirmed that the best average OSGQT fidelity outperformed the best average SGQT fidelity (see supplementary info). Similarly, OSGQT continues to outperform  SGQT in different statistical noise regimes, see supplementary info.




\emph{Conclusions}:--This work shows a mathematical equivalence between self-guided quantum tomography and single-pixel imaging when the distance measure between the unknown ``state'' (or image) and the estimate state are set to be linear, thereby introducing the concept of self-guided imaging.  
It follows that present implementations of single-pixel imaging algorithms can be seen as a subset of iterative self-guided tomography algorithms, where the key feature is that the experiment guides itself to its solution. Having noted this similarity, we take inspiration from OGI, which has been shown to lead to faster convergence in SPI with non-orthogonal measurements, and apply a computational correction to SGQT, introducing OSGQT. We show numerically and experimentally that OSGQT not only converges faster but also achieves higher final fidelities
($95.2\% \rightarrow 99.17\%$ and $92.1\% \rightarrow 95.3\%$ respectively) than SGQT.  This work offers a new conceptual framework to advance SPI and SGQT beyond this paradigm.  

\acknowledgments
We acknowledge financial support from the Engineering and Physical 
Sciences Research Council (EP/Z533166/1 and EP/Z533178/1), the Royal Society (RSRP/R1/211013), the Wits MIND Institute, and the South African Quantum Technology Initiative (SAQuTI).


\begin{thebibliography}{26}%
\makeatletter
\providecommand \@ifxundefined [1]{%
 \@ifx{#1\undefined}
}%
\providecommand \@ifnum [1]{%
 \ifnum #1\expandafter \@firstoftwo
 \else \expandafter \@secondoftwo
 \fi
}%
\providecommand \@ifx [1]{%
 \ifx #1\expandafter \@firstoftwo
 \else \expandafter \@secondoftwo
 \fi
}%
\providecommand \natexlab [1]{#1}%
\providecommand \enquote  [1]{``#1''}%
\providecommand \bibnamefont  [1]{#1}%
\providecommand \bibfnamefont [1]{#1}%
\providecommand \citenamefont [1]{#1}%
\providecommand \href@noop [0]{\@secondoftwo}%
\providecommand \href [0]{\begingroup \@sanitize@url \@href}%
\providecommand \@href[1]{\@@startlink{#1}\@@href}%
\providecommand \@@href[1]{\endgroup#1\@@endlink}%
\providecommand \@sanitize@url [0]{\catcode `\\12\catcode `\$12\catcode
  `\&12\catcode `\#12\catcode `\^12\catcode `\_12\catcode `\%12\relax}%
\providecommand \@@startlink[1]{}%
\providecommand \@@endlink[0]{}%
\providecommand \url  [0]{\begingroup\@sanitize@url \@url }%
\providecommand \@url [1]{\endgroup\@href {#1}{\urlprefix }}%
\providecommand \urlprefix  [0]{URL }%
\providecommand \Eprint [0]{\href }%
\providecommand \doibase [0]{https://doi.org/}%
\providecommand \selectlanguage [0]{\@gobble}%
\providecommand \bibinfo  [0]{\@secondoftwo}%
\providecommand \bibfield  [0]{\@secondoftwo}%
\providecommand \translation [1]{[#1]}%
\providecommand \BibitemOpen [0]{}%
\providecommand \bibitemStop [0]{}%
\providecommand \bibitemNoStop [0]{.\EOS\space}%
\providecommand \EOS [0]{\spacefactor3000\relax}%
\providecommand \BibitemShut  [1]{\csname bibitem#1\endcsname}%
\let\auto@bib@innerbib\@empty
\bibitem [{\citenamefont {Gebhart}\ \emph {et~al.}(2023)\citenamefont
  {Gebhart}, \citenamefont {Santagati}, \citenamefont {Gentile}, \citenamefont
  {Gauger}, \citenamefont {Craig}, \citenamefont {Ares}, \citenamefont
  {Banchi}, \citenamefont {Marquardt}, \citenamefont {Pezz{\`e}},\ and\
  \citenamefont {Bonato}}]{Gebhart}%
  \BibitemOpen
  \bibfield  {author} {\bibinfo {author} {\bibfnamefont {V.}~\bibnamefont
  {Gebhart}}, \bibinfo {author} {\bibfnamefont {R.}~\bibnamefont {Santagati}},
  \bibinfo {author} {\bibfnamefont {A.~A.}\ \bibnamefont {Gentile}}, \bibinfo
  {author} {\bibfnamefont {E.~M.}\ \bibnamefont {Gauger}}, \bibinfo {author}
  {\bibfnamefont {D.}~\bibnamefont {Craig}}, \bibinfo {author} {\bibfnamefont
  {N.}~\bibnamefont {Ares}}, \bibinfo {author} {\bibfnamefont {L.}~\bibnamefont
  {Banchi}}, \bibinfo {author} {\bibfnamefont {F.}~\bibnamefont {Marquardt}},
  \bibinfo {author} {\bibfnamefont {L.}~\bibnamefont {Pezz{\`e}}},\ and\
  \bibinfo {author} {\bibfnamefont {C.}~\bibnamefont {Bonato}},\ }\bibfield
  {title} {\bibinfo {title} {Learning quantum systems},\ }\href
  {https://doi.org/10.1038/s42254-022-00552-1} {\bibfield  {journal} {\bibinfo
  {journal} {Nature Reviews Physics}\ }\textbf {\bibinfo {volume} {5}},\
  \bibinfo {pages} {141} (\bibinfo {year} {2023})}\BibitemShut {NoStop}%
\bibitem [{\citenamefont {Gibson}\ \emph {et~al.}(2020)\citenamefont {Gibson},
  \citenamefont {Johnson},\ and\ \citenamefont {Padgett}}]{Gibson:20}%
  \BibitemOpen
  \bibfield  {author} {\bibinfo {author} {\bibfnamefont {G.~M.}\ \bibnamefont
  {Gibson}}, \bibinfo {author} {\bibfnamefont {S.~D.}\ \bibnamefont
  {Johnson}},\ and\ \bibinfo {author} {\bibfnamefont {M.~J.}\ \bibnamefont
  {Padgett}},\ }\bibfield  {title} {\bibinfo {title} {Single-pixel imaging 12
  years on: a review},\ }\href {https://doi.org/10.1364/OE.403195} {\bibfield
  {journal} {\bibinfo  {journal} {Opt. Express}\ }\textbf {\bibinfo {volume}
  {28}},\ \bibinfo {pages} {28190} (\bibinfo {year} {2020})}\BibitemShut
  {NoStop}%
\bibitem [{\citenamefont {Mauro~D’Ariano}\ \emph {et~al.}(2003)\citenamefont
  {Mauro~D’Ariano}, \citenamefont {Paris},\ and\ \citenamefont
  {Sacchi}}]{MAURODARIANO2003205}%
  \BibitemOpen
  \bibfield  {author} {\bibinfo {author} {\bibfnamefont {G.}~\bibnamefont
  {Mauro~D’Ariano}}, \bibinfo {author} {\bibfnamefont {M.~G.}\ \bibnamefont
  {Paris}},\ and\ \bibinfo {author} {\bibfnamefont {M.~F.}\ \bibnamefont
  {Sacchi}},\ }\bibfield  {title} {\bibinfo {title} {Quantum {Tomography}},\
  }in\ \href {https://doi.org/10.1016/S1076-5670(03)80065-4} {\emph {\bibinfo
  {booktitle} {Advances in {Imaging} and {Electron} {Physics}}}},\ Vol.\
  \bibinfo {volume} {128}\ (\bibinfo  {publisher} {Elsevier},\ \bibinfo {year}
  {2003})\ pp.\ \bibinfo {pages} {205--308}\BibitemShut {NoStop}%
\bibitem [{\citenamefont {Bolduc}\ \emph {et~al.}(2017)\citenamefont {Bolduc},
  \citenamefont {Knee}, \citenamefont {Gauger},\ and\ \citenamefont
  {Leach}}]{RN1363}%
  \BibitemOpen
  \bibfield  {author} {\bibinfo {author} {\bibfnamefont {E.}~\bibnamefont
  {Bolduc}}, \bibinfo {author} {\bibfnamefont {G.~C.}\ \bibnamefont {Knee}},
  \bibinfo {author} {\bibfnamefont {E.~M.}\ \bibnamefont {Gauger}},\ and\
  \bibinfo {author} {\bibfnamefont {J.}~\bibnamefont {Leach}},\ }\bibfield
  {title} {\bibinfo {title} {Projected gradient descent algorithms for quantum
  state tomography},\ }\bibfield  {journal} {\bibinfo  {journal} {Npj Quantum
  Information}\ }\textbf {\bibinfo {volume} {3}},\ \href
  {https://doi.org/10.1038/s41534-017-0043-1} {10.1038/s41534-017-0043-1}
  (\bibinfo {year} {2017})\BibitemShut {NoStop}%
\bibitem [{\citenamefont {James}\ \emph {et~al.}(2001)\citenamefont {James},
  \citenamefont {Kwiat}, \citenamefont {Munro},\ and\ \citenamefont
  {White}}]{PhysRevA.64.052312}%
  \BibitemOpen
  \bibfield  {author} {\bibinfo {author} {\bibfnamefont {D.~F.~V.}\
  \bibnamefont {James}}, \bibinfo {author} {\bibfnamefont {P.~G.}\ \bibnamefont
  {Kwiat}}, \bibinfo {author} {\bibfnamefont {W.~J.}\ \bibnamefont {Munro}},\
  and\ \bibinfo {author} {\bibfnamefont {A.~G.}\ \bibnamefont {White}},\
  }\bibfield  {title} {\bibinfo {title} {Measurement of qubits},\ }\href
  {https://doi.org/10.1103/PhysRevA.64.052312} {\bibfield  {journal} {\bibinfo
  {journal} {Phys. Rev. A}\ }\textbf {\bibinfo {volume} {64}},\ \bibinfo
  {pages} {052312} (\bibinfo {year} {2001})}\BibitemShut {NoStop}%
\bibitem [{\citenamefont {Agnew}\ \emph {et~al.}(2011)\citenamefont {Agnew},
  \citenamefont {Leach}, \citenamefont {McLaren}, \citenamefont {Roux},\ and\
  \citenamefont {Boyd}}]{RN1365}%
  \BibitemOpen
  \bibfield  {author} {\bibinfo {author} {\bibfnamefont {M.}~\bibnamefont
  {Agnew}}, \bibinfo {author} {\bibfnamefont {J.}~\bibnamefont {Leach}},
  \bibinfo {author} {\bibfnamefont {M.}~\bibnamefont {McLaren}}, \bibinfo
  {author} {\bibfnamefont {F.~S.}\ \bibnamefont {Roux}},\ and\ \bibinfo
  {author} {\bibfnamefont {R.~W.}\ \bibnamefont {Boyd}},\ }\bibfield  {title}
  {\bibinfo {title} {Tomography of the quantum state of photons entangled in
  high dimensions},\ }\href {https://doi.org/10.1103/PhysRevA.84.062101}
  {\bibfield  {journal} {\bibinfo  {journal} {Phys. Rev. A}\ }\textbf {\bibinfo
  {volume} {84}},\ \bibinfo {pages} {062101} (\bibinfo {year}
  {2011})}\BibitemShut {NoStop}%
\bibitem [{\citenamefont {Ferrie}(2014)}]{RN1358}%
  \BibitemOpen
  \bibfield  {author} {\bibinfo {author} {\bibfnamefont {C.}~\bibnamefont
  {Ferrie}},\ }\bibfield  {title} {\bibinfo {title} {Self-guided quantum
  tomography},\ }\href {https://doi.org/10.1103/PhysRevLett.113.190404}
  {\bibfield  {journal} {\bibinfo  {journal} {Phys. Rev. Lett.}\ }\textbf
  {\bibinfo {volume} {113}},\ \bibinfo {pages} {190404} (\bibinfo {year}
  {2014})}\BibitemShut {NoStop}%
\bibitem [{\citenamefont {Thew}\ \emph {et~al.}(2002)\citenamefont {Thew},
  \citenamefont {Nemoto}, \citenamefont {White},\ and\ \citenamefont
  {Munro}}]{thew_qudit_2002}%
  \BibitemOpen
  \bibfield  {author} {\bibinfo {author} {\bibfnamefont {R.~T.}\ \bibnamefont
  {Thew}}, \bibinfo {author} {\bibfnamefont {K.}~\bibnamefont {Nemoto}},
  \bibinfo {author} {\bibfnamefont {A.~G.}\ \bibnamefont {White}},\ and\
  \bibinfo {author} {\bibfnamefont {W.~J.}\ \bibnamefont {Munro}},\ }\bibfield
  {title} {\bibinfo {title} {Qudit quantum-state tomography},\ }\href
  {https://doi.org/10.1103/PhysRevA.66.012303} {\bibfield  {journal} {\bibinfo
  {journal} {Physical Review A}\ }\textbf {\bibinfo {volume} {66}},\ \bibinfo
  {pages} {012303} (\bibinfo {year} {2002})}\BibitemShut {NoStop}%
\bibitem [{\citenamefont {Chapman}\ \emph {et~al.}(2016)\citenamefont
  {Chapman}, \citenamefont {Ferrie},\ and\ \citenamefont
  {Peruzzo}}]{PhysRevLett.117.040402}%
  \BibitemOpen
  \bibfield  {author} {\bibinfo {author} {\bibfnamefont {R.~J.}\ \bibnamefont
  {Chapman}}, \bibinfo {author} {\bibfnamefont {C.}~\bibnamefont {Ferrie}},\
  and\ \bibinfo {author} {\bibfnamefont {A.}~\bibnamefont {Peruzzo}},\
  }\bibfield  {title} {\bibinfo {title} {Experimental demonstration of
  self-guided quantum tomography},\ }\href
  {https://doi.org/10.1103/PhysRevLett.117.040402} {\bibfield  {journal}
  {\bibinfo  {journal} {Phys. Rev. Lett.}\ }\textbf {\bibinfo {volume} {117}},\
  \bibinfo {pages} {040402} (\bibinfo {year} {2016})}\BibitemShut {NoStop}%
\bibitem [{\citenamefont {Rambach}\ \emph {et~al.}(2021)\citenamefont
  {Rambach}, \citenamefont {Qaryan}, \citenamefont {Kewming}, \citenamefont
  {Ferrie}, \citenamefont {White},\ and\ \citenamefont
  {Romero}}]{Rambach_tomography_2021}%
  \BibitemOpen
  \bibfield  {author} {\bibinfo {author} {\bibfnamefont {M.}~\bibnamefont
  {Rambach}}, \bibinfo {author} {\bibfnamefont {M.}~\bibnamefont {Qaryan}},
  \bibinfo {author} {\bibfnamefont {M.}~\bibnamefont {Kewming}}, \bibinfo
  {author} {\bibfnamefont {C.}~\bibnamefont {Ferrie}}, \bibinfo {author}
  {\bibfnamefont {A.~G.}\ \bibnamefont {White}},\ and\ \bibinfo {author}
  {\bibfnamefont {J.}~\bibnamefont {Romero}},\ }\bibfield  {title} {\bibinfo
  {title} {Robust and efficient high-dimensional quantum state tomography},\
  }\href {https://doi.org/10.1103/PhysRevLett.126.100402} {\bibfield  {journal}
  {\bibinfo  {journal} {Phys. Rev. Lett.}\ }\textbf {\bibinfo {volume} {126}},\
  \bibinfo {pages} {100402} (\bibinfo {year} {2021})}\BibitemShut {NoStop}%
\bibitem [{\citenamefont {Serino}\ \emph {et~al.}(2025)\citenamefont {Serino},
  \citenamefont {Rambach}, \citenamefont {Brecht}, \citenamefont {Romero},\
  and\ \citenamefont {Silberhorn}}]{serino_self-guided_2025}%
  \BibitemOpen
  \bibfield  {author} {\bibinfo {author} {\bibfnamefont {L.}~\bibnamefont
  {Serino}}, \bibinfo {author} {\bibfnamefont {M.}~\bibnamefont {Rambach}},
  \bibinfo {author} {\bibfnamefont {B.}~\bibnamefont {Brecht}}, \bibinfo
  {author} {\bibfnamefont {J.}~\bibnamefont {Romero}},\ and\ \bibinfo {author}
  {\bibfnamefont {C.}~\bibnamefont {Silberhorn}},\ }\bibfield  {title}
  {\bibinfo {title} {Self-guided tomography of time-frequency qudits},\ }\href
  {https://doi.org/10.1088/2058-9565/adb0ea} {\bibfield  {journal} {\bibinfo
  {journal} {Quantum Science and Technology}\ }\textbf {\bibinfo {volume}
  {10}},\ \bibinfo {pages} {025024} (\bibinfo {year} {2025})}\BibitemShut
  {NoStop}%
\bibitem [{\citenamefont {Hou}\ \emph {et~al.}(2020)\citenamefont {Hou},
  \citenamefont {Tang}, \citenamefont {Ferrie}, \citenamefont {Xiang},
  \citenamefont {Li},\ and\ \citenamefont {Guo}}]{RN1361}%
  \BibitemOpen
  \bibfield  {author} {\bibinfo {author} {\bibfnamefont {Z.}~\bibnamefont
  {Hou}}, \bibinfo {author} {\bibfnamefont {J.-F.}\ \bibnamefont {Tang}},
  \bibinfo {author} {\bibfnamefont {C.}~\bibnamefont {Ferrie}}, \bibinfo
  {author} {\bibfnamefont {G.-Y.}\ \bibnamefont {Xiang}}, \bibinfo {author}
  {\bibfnamefont {C.-F.}\ \bibnamefont {Li}},\ and\ \bibinfo {author}
  {\bibfnamefont {G.-C.}\ \bibnamefont {Guo}},\ }\bibfield  {title} {\bibinfo
  {title} {Experimental realization of self-guided quantum process
  tomography},\ }\href {https://doi.org/10.1103/PhysRevA.101.022317} {\bibfield
   {journal} {\bibinfo  {journal} {Phys. Rev. A}\ }\textbf {\bibinfo {volume}
  {101}},\ \bibinfo {pages} {022317} (\bibinfo {year} {2020})}\BibitemShut
  {NoStop}%
\bibitem [{\citenamefont {Duarte}\ \emph {et~al.}(2008)\citenamefont {Duarte},
  \citenamefont {Davenport}, \citenamefont {Takhar}, \citenamefont {Laska},
  \citenamefont {Sun}, \citenamefont {Kelly},\ and\ \citenamefont
  {Baraniuk}}]{4472247}%
  \BibitemOpen
  \bibfield  {author} {\bibinfo {author} {\bibfnamefont {M.~F.}\ \bibnamefont
  {Duarte}}, \bibinfo {author} {\bibfnamefont {M.~A.}\ \bibnamefont
  {Davenport}}, \bibinfo {author} {\bibfnamefont {D.}~\bibnamefont {Takhar}},
  \bibinfo {author} {\bibfnamefont {J.~N.}\ \bibnamefont {Laska}}, \bibinfo
  {author} {\bibfnamefont {T.}~\bibnamefont {Sun}}, \bibinfo {author}
  {\bibfnamefont {K.~F.}\ \bibnamefont {Kelly}},\ and\ \bibinfo {author}
  {\bibfnamefont {R.~G.}\ \bibnamefont {Baraniuk}},\ }\bibfield  {title}
  {\bibinfo {title} {Single-pixel imaging via compressive sampling},\ }\href
  {https://doi.org/10.1109/MSP.2007.914730} {\bibfield  {journal} {\bibinfo
  {journal} {IEEE Signal Processing Magazine}\ }\textbf {\bibinfo {volume}
  {25}},\ \bibinfo {pages} {83} (\bibinfo {year} {2008})}\BibitemShut {NoStop}%
\bibitem [{\citenamefont {Kallepalli}\ \emph {et~al.}(2022)\citenamefont
  {Kallepalli}, \citenamefont {Viani}, \citenamefont {Stellinga}, \citenamefont
  {Rotunno}, \citenamefont {Bowman}, \citenamefont {Gibson}, \citenamefont
  {Sun}, \citenamefont {Rosi}, \citenamefont {Frabboni}, \citenamefont
  {Balboni}, \citenamefont {Migliori}, \citenamefont {Grillo},\ and\
  \citenamefont {Padgett}}]{kallepalli_challenging_2022}%
  \BibitemOpen
  \bibfield  {author} {\bibinfo {author} {\bibfnamefont {A.}~\bibnamefont
  {Kallepalli}}, \bibinfo {author} {\bibfnamefont {L.}~\bibnamefont {Viani}},
  \bibinfo {author} {\bibfnamefont {D.}~\bibnamefont {Stellinga}}, \bibinfo
  {author} {\bibfnamefont {E.}~\bibnamefont {Rotunno}}, \bibinfo {author}
  {\bibfnamefont {R.}~\bibnamefont {Bowman}}, \bibinfo {author} {\bibfnamefont
  {G.~M.}\ \bibnamefont {Gibson}}, \bibinfo {author} {\bibfnamefont {M.-J.}\
  \bibnamefont {Sun}}, \bibinfo {author} {\bibfnamefont {P.}~\bibnamefont
  {Rosi}}, \bibinfo {author} {\bibfnamefont {S.}~\bibnamefont {Frabboni}},
  \bibinfo {author} {\bibfnamefont {R.}~\bibnamefont {Balboni}}, \bibinfo
  {author} {\bibfnamefont {A.}~\bibnamefont {Migliori}}, \bibinfo {author}
  {\bibfnamefont {V.}~\bibnamefont {Grillo}},\ and\ \bibinfo {author}
  {\bibfnamefont {M.~J.}\ \bibnamefont {Padgett}},\ }\bibfield  {title}
  {\bibinfo {title} {Challenging {Point} {Scanning} across {Electron}
  {Microscopy} and {Optical} {Imaging} using {Computational} {Imaging}},\
  }\href {https://doi.org/10.34133/icomputing.0001} {\bibfield  {journal}
  {\bibinfo  {journal} {Intelligent Computing}\ }\textbf {\bibinfo {volume}
  {2022}},\ \bibinfo {pages} {0001} (\bibinfo {year} {2022})}\BibitemShut
  {NoStop}%
\bibitem [{\citenamefont {Popa}\ and\ \citenamefont
  {Zdunek}(2004)}]{POPA2004579}%
  \BibitemOpen
  \bibfield  {author} {\bibinfo {author} {\bibfnamefont {C.}~\bibnamefont
  {Popa}}\ and\ \bibinfo {author} {\bibfnamefont {R.}~\bibnamefont {Zdunek}},\
  }\bibfield  {title} {\bibinfo {title} {Kaczmarz extended algorithm for
  tomographic image reconstruction from limited-data},\ }\href
  {https://doi.org/https://doi.org/10.1016/j.matcom.2004.01.021} {\bibfield
  {journal} {\bibinfo  {journal} {Mathematics and Computers in Simulation}\
  }\textbf {\bibinfo {volume} {65}},\ \bibinfo {pages} {579} (\bibinfo {year}
  {2004})}\BibitemShut {NoStop}%
\bibitem [{\citenamefont {Strohmer}\ and\ \citenamefont
  {Vershynin}(2009)}]{strohmer2009randomized}%
  \BibitemOpen
  \bibfield  {author} {\bibinfo {author} {\bibfnamefont {T.}~\bibnamefont
  {Strohmer}}\ and\ \bibinfo {author} {\bibfnamefont {R.}~\bibnamefont
  {Vershynin}},\ }\bibfield  {title} {\bibinfo {title} {A randomized kaczmarz
  algorithm with exponential convergence},\ }\href
  {https://doi.org/10.1007/s00041-008-9030-4} {\bibfield  {journal} {\bibinfo
  {journal} {Journal of Fourier Analysis and Applications}\ }\textbf {\bibinfo
  {volume} {15}},\ \bibinfo {pages} {262} (\bibinfo {year} {2009})}\BibitemShut
  {NoStop}%
\bibitem [{\citenamefont {Pires}\ \emph {et~al.}(2016)\citenamefont {Pires},
  \citenamefont {Pereira}, \citenamefont {Pereira}, \citenamefont {Mansano},\
  and\ \citenamefont {Papa}}]{pires2016projections}%
  \BibitemOpen
  \bibfield  {author} {\bibinfo {author} {\bibfnamefont {R.~G.}\ \bibnamefont
  {Pires}}, \bibinfo {author} {\bibfnamefont {D.~R.}\ \bibnamefont {Pereira}},
  \bibinfo {author} {\bibfnamefont {L.~A.}\ \bibnamefont {Pereira}}, \bibinfo
  {author} {\bibfnamefont {A.~F.}\ \bibnamefont {Mansano}},\ and\ \bibinfo
  {author} {\bibfnamefont {J.~P.}\ \bibnamefont {Papa}},\ }\bibfield  {title}
  {\bibinfo {title} {Projections onto convex sets parameter estimation through
  harmony search and its application for image restoration},\ }\href
  {https://doi.org/10.1007/s11047-015-9507-4} {\bibfield  {journal} {\bibinfo
  {journal} {Natural Computing}\ }\textbf {\bibinfo {volume} {15}},\ \bibinfo
  {pages} {493} (\bibinfo {year} {2016})}\BibitemShut {NoStop}%
\bibitem [{\citenamefont {Sun}\ \emph {et~al.}(2020)\citenamefont {Sun},
  \citenamefont {Gu},\ and\ \citenamefont {Tang}}]{sun2020randomized}%
  \BibitemOpen
  \bibfield  {author} {\bibinfo {author} {\bibfnamefont {M.-L.}\ \bibnamefont
  {Sun}}, \bibinfo {author} {\bibfnamefont {C.-Q.}\ \bibnamefont {Gu}},\ and\
  \bibinfo {author} {\bibfnamefont {P.-F.}\ \bibnamefont {Tang}},\ }\bibfield
  {title} {\bibinfo {title} {On randomized sampling kaczmarz method with
  application in compressed sensing},\ }\href
  {https://doi.org/10.1155/2020/7464212} {\bibfield  {journal} {\bibinfo
  {journal} {Mathematical Problems in Engineering}\ }\textbf {\bibinfo {volume}
  {2020}},\ \bibinfo {pages} {7464212} (\bibinfo {year} {2020})}\BibitemShut
  {NoStop}%
\bibitem [{\citenamefont {Spall}(1992)}]{RN1367}%
  \BibitemOpen
  \bibfield  {author} {\bibinfo {author} {\bibfnamefont {J.~C.}\ \bibnamefont
  {Spall}},\ }\bibfield  {title} {\bibinfo {title} {Multivariate
  stochastic-approximation using a simultaneous perturbation gradient
  approximation},\ }\href {https://doi.org/Doi 10.1109/9.119632} {\bibfield
  {journal} {\bibinfo  {journal} {Ieee Transactions on Automatic Control}\
  }\textbf {\bibinfo {volume} {37}},\ \bibinfo {pages} {332} (\bibinfo {year}
  {1992})}\BibitemShut {NoStop}%
\bibitem [{\citenamefont {Yu}\ \emph {et~al.}(2021)\citenamefont {Yu},
  \citenamefont {Zhu}, \citenamefont {Li}, \citenamefont {Wang},\ and\
  \citenamefont {Cao}}]{Yu_gradient-descent_2021}%
  \BibitemOpen
  \bibfield  {author} {\bibinfo {author} {\bibfnamefont {W.-K.}\ \bibnamefont
  {Yu}}, \bibinfo {author} {\bibfnamefont {C.-X.}\ \bibnamefont {Zhu}},
  \bibinfo {author} {\bibfnamefont {Y.-X.}\ \bibnamefont {Li}}, \bibinfo
  {author} {\bibfnamefont {S.-F.}\ \bibnamefont {Wang}},\ and\ \bibinfo
  {author} {\bibfnamefont {C.}~\bibnamefont {Cao}},\ }\bibfield  {title}
  {\bibinfo {title} {Gradient-descent-like ghost imaging},\ }\href
  {https://doi.org/10.3390/s21227559} {\bibfield  {journal} {\bibinfo
  {journal} {Sensors}\ }\textbf {\bibinfo {volume} {21}},\ \bibinfo {pages}
  {7559} (\bibinfo {year} {2021})}\BibitemShut {NoStop}%
\bibitem [{\citenamefont {Pittman}\ \emph {et~al.}(1995)\citenamefont
  {Pittman}, \citenamefont {Shih}, \citenamefont {Strekalov},\ and\
  \citenamefont {Sergienko}}]{pittman_optical_1995}%
  \BibitemOpen
  \bibfield  {author} {\bibinfo {author} {\bibfnamefont {T.~B.}\ \bibnamefont
  {Pittman}}, \bibinfo {author} {\bibfnamefont {Y.~H.}\ \bibnamefont {Shih}},
  \bibinfo {author} {\bibfnamefont {D.~V.}\ \bibnamefont {Strekalov}},\ and\
  \bibinfo {author} {\bibfnamefont {A.~V.}\ \bibnamefont {Sergienko}},\
  }\bibfield  {title} {\bibinfo {title} {Optical imaging by means of two-photon
  quantum entanglement},\ }\href {https://doi.org/10.1103/PhysRevA.52.R3429}
  {\bibfield  {journal} {\bibinfo  {journal} {Physical Review A}\ }\textbf
  {\bibinfo {volume} {52}},\ \bibinfo {pages} {R3429} (\bibinfo {year}
  {1995})}\BibitemShut {NoStop}%
\bibitem [{\citenamefont {Bennink}\ \emph {et~al.}(2002)\citenamefont
  {Bennink}, \citenamefont {Bentley},\ and\ \citenamefont
  {Boyd}}]{bennink_two_photon_2002}%
  \BibitemOpen
  \bibfield  {author} {\bibinfo {author} {\bibfnamefont {R.~S.}\ \bibnamefont
  {Bennink}}, \bibinfo {author} {\bibfnamefont {S.~J.}\ \bibnamefont
  {Bentley}},\ and\ \bibinfo {author} {\bibfnamefont {R.~W.}\ \bibnamefont
  {Boyd}},\ }\bibfield  {title} {\bibinfo {title} {``two-photon'' coincidence
  imaging with a classical source},\ }\href
  {https://doi.org/10.1103/PhysRevLett.89.113601} {\bibfield  {journal}
  {\bibinfo  {journal} {Phys. Rev. Lett.}\ }\textbf {\bibinfo {volume} {89}},\
  \bibinfo {pages} {113601} (\bibinfo {year} {2002})}\BibitemShut {NoStop}%
\bibitem [{\citenamefont {Erkmen}\ and\ \citenamefont
  {Shapiro}(2010)}]{Erkmen:10}%
  \BibitemOpen
  \bibfield  {author} {\bibinfo {author} {\bibfnamefont {B.~I.}\ \bibnamefont
  {Erkmen}}\ and\ \bibinfo {author} {\bibfnamefont {J.~H.}\ \bibnamefont
  {Shapiro}},\ }\bibfield  {title} {\bibinfo {title} {Ghost imaging: from
  quantum to classical to computational},\ }\href
  {https://doi.org/10.1364/AOP.2.000405} {\bibfield  {journal} {\bibinfo
  {journal} {Adv. Opt. Photon.}\ }\textbf {\bibinfo {volume} {2}},\ \bibinfo
  {pages} {405} (\bibinfo {year} {2010})}\BibitemShut {NoStop}%
\bibitem [{\citenamefont {Yao}\ and\ \citenamefont
  {Padgett}(2011)}]{yao2011orbital}%
  \BibitemOpen
  \bibfield  {author} {\bibinfo {author} {\bibfnamefont {A.~M.}\ \bibnamefont
  {Yao}}\ and\ \bibinfo {author} {\bibfnamefont {M.~J.}\ \bibnamefont
  {Padgett}},\ }\bibfield  {title} {\bibinfo {title} {Orbital angular momentum:
  origins, behavior and applications},\ }\href
  {https://doi.org/10.1364/AOP.3.000161} {\bibfield  {journal} {\bibinfo
  {journal} {Advances in Optics and Photonics}\ }\textbf {\bibinfo {volume}
  {3}},\ \bibinfo {pages} {161} (\bibinfo {year} {2011})}\BibitemShut {NoStop}%
\bibitem [{\citenamefont {Allen}\ \emph {et~al.}(2016)\citenamefont {Allen},
  \citenamefont {Barnett},\ and\ \citenamefont {Padgett}}]{allen2016optical}%
  \BibitemOpen
  \bibfield  {author} {\bibinfo {author} {\bibfnamefont {L.}~\bibnamefont
  {Allen}}, \bibinfo {author} {\bibfnamefont {S.~M.}\ \bibnamefont {Barnett}},\
  and\ \bibinfo {author} {\bibfnamefont {M.~J.}\ \bibnamefont {Padgett}},\
  }\href@noop {} {\emph {\bibinfo {title} {Optical angular momentum}}}\
  (\bibinfo  {publisher} {CRC press},\ \bibinfo {year} {2016})\BibitemShut
  {NoStop}%
\bibitem [{\citenamefont {Mair}\ \emph {et~al.}(2001)\citenamefont {Mair},
  \citenamefont {Vaziri}, \citenamefont {Weihs},\ and\ \citenamefont
  {Zeilinger}}]{mair2001entanglement}%
  \BibitemOpen
  \bibfield  {author} {\bibinfo {author} {\bibfnamefont {A.}~\bibnamefont
  {Mair}}, \bibinfo {author} {\bibfnamefont {A.}~\bibnamefont {Vaziri}},
  \bibinfo {author} {\bibfnamefont {G.}~\bibnamefont {Weihs}},\ and\ \bibinfo
  {author} {\bibfnamefont {A.}~\bibnamefont {Zeilinger}},\ }\bibfield  {title}
  {\bibinfo {title} {Entanglement of the orbital angular momentum states of
  photons},\ }\href {https://doi.org/10.1038/35085529} {\bibfield  {journal}
  {\bibinfo  {journal} {Nature}\ }\textbf {\bibinfo {volume} {412}},\ \bibinfo
  {pages} {313} (\bibinfo {year} {2001})}\BibitemShut {NoStop}%
\end{thebibliography}
\end{document}


\title{
Supplementary Information for Orthogonalised Self-Guided Quantum Tomography: Insights from Single-Pixel Imaging}

\author{Kiki Dekkers$^{1*}$, Alice Ruget$^{1*}$, Fazilah Nothlawala$^{2}$, Sabrina Henry$^{1}$, Stirling Scholes$^{1}$, Miles Padgett$^{3}$, Andrew Forbes$^{2}$, Isaac Nape$^{2}$, and Jonathan Leach$^{1,\, \dagger}$}

\affiliation{$^{1}$School of Engineering and Physical Sciences, Heriot-Watt University, Edinburgh, EH14 4AS, UK}
\affiliation{$^{2}$School of Physics, University of the Witwatersrand, Private Bag 3, Wits 2050, South Africa}
\affiliation{$^{3}$University of Glasgow, Glasgow, G12 8QQ, UK}

\thanks{These authors contributed equally}
\email{$^{\dagger}$ j.leach@hw.ac.uk}
















\maketitle


 \begin{figure*}
    \centering
    \includegraphics[width=1\linewidth]{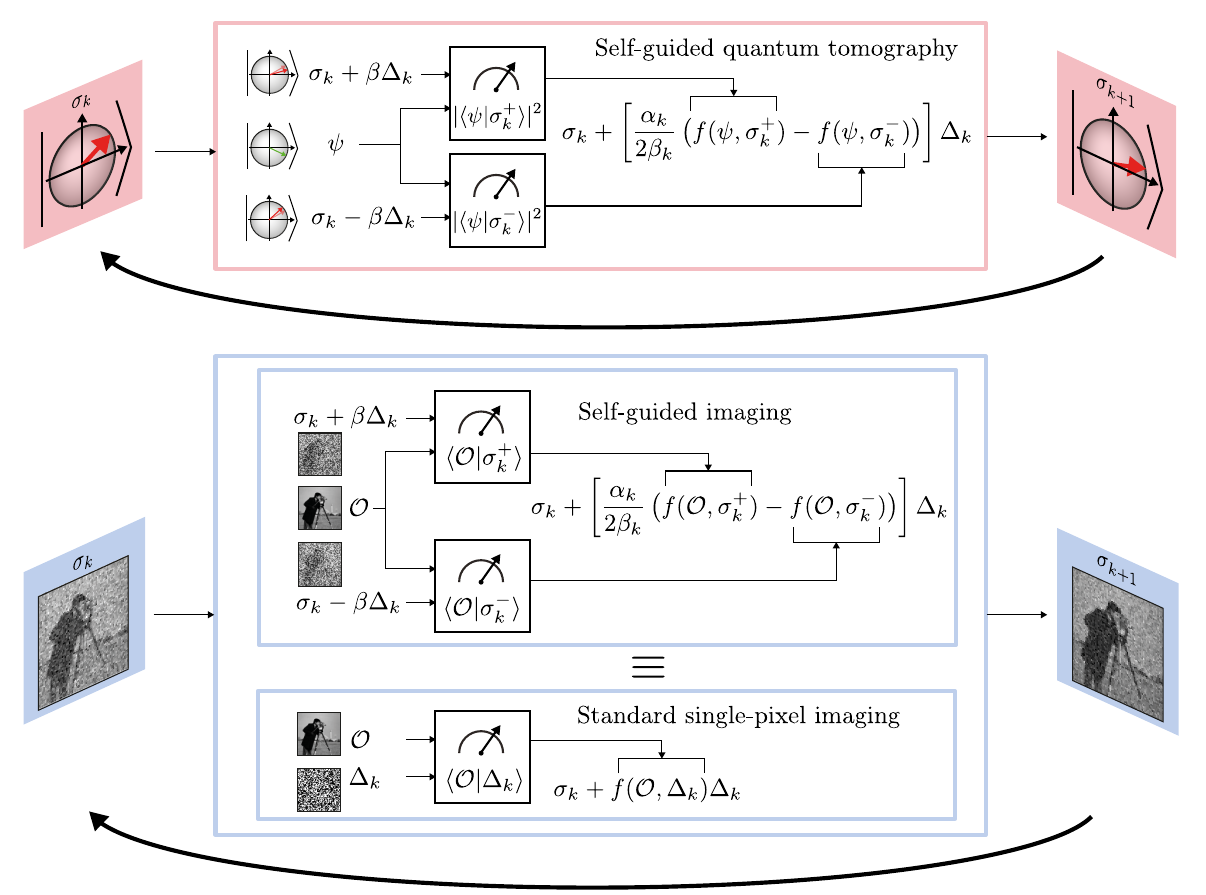}
    \caption{Illustration of self-guided quantum tomography, self-guided imaging and single-pixel imaging. We introduce self-guided imaging as self-guided quantum tomography applied to imaging, i.e. having a linear distance measure. The self-guided imaging algorithm is equivalent to standard single-pixel imaging.  Each of these approaches is a stochastic gradient descent algorithm.}
    \label{fig:SPI_vs_SelfGuiding_concept}
\end{figure*}

\section{Background} \label{S: background}
\vspace{-5pt}
\subsection{Quantum Tomography}
\label{S: quantum tomography}
\vspace{-10pt}
Quantum tomography is the process of establishing an unknown quantum state $\rho$ from an informationally complete set of known measurements $\{A_k\}$ and the corresponding outcomes $\{p_k\}$\cite{MAURODARIANO2003205}.   The probability that the data is observed given a measurement is given by the Born rule $p_k = \rm{Tr}( \rho \hat{A}_k )$, which for pure states $|\psi\rangle$ simplifies to the standard complex overlap squared of the measurement and the state $p_k =  |\langle A_k | \psi\rangle|^2 $.  Many approaches to efficiently collecting the data and then optimally solving the inverse problem have been developed \cite{RN1363}.  This research is required as the quantum tomography problem scales poorly with the dimension of the state \cite{PhysRevA.64.052312, RN1365}, requiring of order $\mathcal{O}(d^2)$ individual measurement outcomes for $d$-dimensional states.  The inversion of this large dataset can be computationally intensive and therefore time consuming.  Quantum process tomography, where quantum channels are characterised, scales even more poorly with of order $\mathcal{O}(d^4)$ outcomes required. 

\subsection{Optimisation strategies within the framework of gradient descent} \label{S: optimisation strategies}
\vspace{-10pt}
A broad class of these optimisation strategies can be understood within the framework of gradient descent, where an estimate of the underlying state or image is iteratively refined to minimise a cost function. In this formalism, an update to the current estimate $\sigma_k$ is iteratively updated according to
\begin{align}
\sigma_{k+1} = \sigma_k - \eta \nabla f(\sigma_k),
\end{align}
where $f(\sigma_k)$ is an objective function or distance measure, $\nabla f(\sigma_k)$ its gradient, and $\eta$ a parameter that controls the step size of each iteration. In the case that the form of the gradient is unavailable or expensive to compute, stochastic methods can approximate the gradient through random sampling or perturbation of the control parameters. The simultaneous perturbation stochastic approximation (SPSA) algorithm \cite{RN1367} provides an efficient estimate of the gradient, requiring only two evaluations of the objective function per iteration.  This stochastic approach to calculating the gradient forms the basis of self-guided quantum tomography  (SGQT) that was introduced by Ferrie \cite{RN1358}.

\subsection{Standard single-pixel imaging} \label{S: S}
\vspace{-10pt}
It is now recognised that the maths associated with ghost imaging has a large overlap with that of SPI \cite{Gibson:20}.  We briefly present the standard approaches to both problems to give further context to our work. Ghost imaging is the process of establishing an unknown object from the use of correlated optical fields. The standard solution is found by measuring a series of overlaps of an unknown object with known intensity or illuminating patterns.   

Using the notation from SGQT, the unknown object is $\mathcal{O}$, the intensity pattern is $\Delta$, and $f(\mathcal{O}, \Delta_k) = \langle \mathcal{O} | \Delta_k\rangle$. Note that $\mathcal{O}$ and $\Delta_k$ are both real-valued for SPI, in contrast to SGQT where they can have complex elements.   Importantly, $f(\mathcal{O}, \Delta_k) $ has a linear dependence on $\Delta_k$.  This is in contrast to the case of quantum measurements where experimental feedback, i.e.~fidelity, varies non-linearly as a function of the input state.  An estimate $\sigma$ of the object can be calculated via
\begin{align}\label{GI}
\sigma =  \langle(f(\mathcal{O}, \Delta_k) - \langle f(\mathcal{O}, \Delta) \rangle) (\Delta_k - \langle \Delta \rangle) \rangle.
\end{align}
Here $\langle f(\mathcal{O}, \Delta) \rangle$ is the average overlap and is equal to $\sum_k^N f(\mathcal{O}, \Delta_k)/N$, where $N$ is the total number of measurements.  The average mask is calculated similarly.  Note that $\mathcal{O}$ and $\sigma$ represent images in this case, rather than quantum states in SGQT.

In SPI, there are typically two modes of operation: one that uses structured illumination of the object and one that uses structured detection after flood illumination.  In both cases, though, it is typical to use a digital mirror device (DMD) or spatial light modulator (SLM) to control the illumination/detection pattern, and this enables sufficient control such that $\langle f(\mathcal{O}, \Delta) \rangle$ and $\langle \Delta \rangle$ can be equal to zero.  This is achieved by, for example, using equal and opposite DMD masks $\Delta_k^+$ and $\Delta_k^-$ so that the difference in their signals can be used. In such a case, Eq.~\ref{GI} simplifies so that the image can be estimated via 
\begin{align}
\sigma =  \langle f(\mathcal{O}, \Delta_k)  \Delta_k \rangle  = \sum_k^N \frac{f(\mathcal{O}, \Delta_k) \Delta_k}{N}.
\end{align}

It is in fact common to iteratively estimate the image as the experiment progresses, and in this case, after each iteration, the update to the current image $\sigma_k$ is
\begin{align}
\sigma_{k+1}  & =  \sigma_k + f(\mathcal{O}, \Delta_k) \Delta_k \nonumber \\ 
& = \sigma_k + \langle \mathcal{O} | \Delta_k \rangle \Delta_k
\label{SGI}
\end{align}
We can understand this intuitively as the guess image at the next iteration is equal to the current solution plus the overlap of the mask and the object $f(\mathcal{O}, \Delta_k)$ times the mask $\Delta_k$.  We are iteratively updating a guess solution $\sigma_k$ with $\langle \mathcal{O} | \Delta_k \rangle \Delta_k$.   The normalisation of the solution can be achieved at the final iteration, i.e., $\sigma = \sigma_{N}/N$.\\
\\
\clearpage
\newpage
\section{SPI and SGI} \label{S: SPI and SGI}
\vspace{-10pt}
Fig.~\ref{fig:SPI_vs_SelfGuiding_concept} provides an illustration of the equivalence of single-pixel imaging and self-guided imaging.  Each of these approaches is a stochastic gradient descent algorithm where the algorithm guides itself to the solution. 

\subsection{Results for SPI vs SGI for random masks} \label{S: SPI SGI random masks }
\begin{figure}[b]
    \centering
    \includegraphics[]{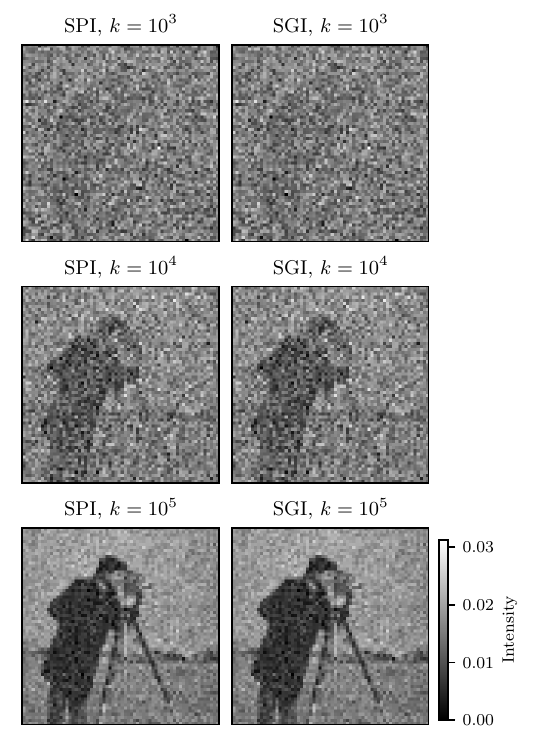}
    \caption{Comparison of output images for single-pixel imaging vs.~self-imaging reconstruction for random masks.  Results are for a 64$\times$64 pixel image.}
    \label{fig:SPI_vs_SelfGuiding}
\end{figure}

\begin{figure}
    \centering
    \includegraphics[]{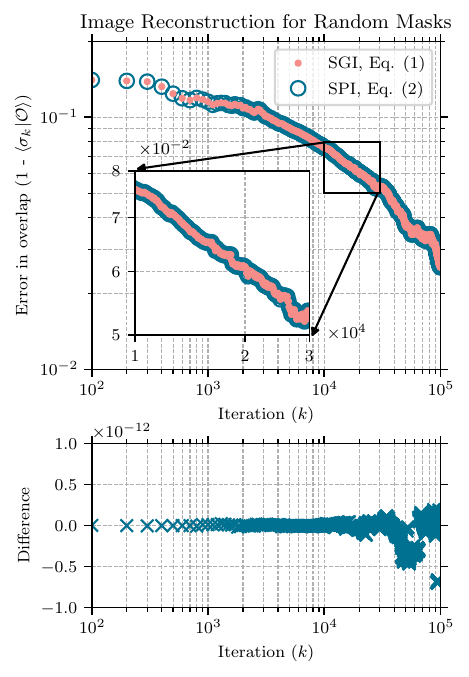}
    \caption{Comparison of error for single-pixel imaging vs.~self-guided imaging for random masks. Only every 100$^{\rm th}$ data point is shown for clarity, and the results are for a 64$\times$64 pixel image.}
    \label{fig:SPI_vs_SelfGuiding_data}
\end{figure}

\vspace{-10pt}
Fig.~\ref{fig:SPI_vs_SelfGuiding} compares the output of single-pixel imaging to that of self-guided imaging for random masks applied to an image of 64$\times$64 pixels.  Each image is normalised such that $\langle \mathcal{O} |\mathcal{O} \rangle = 1$, and each random mask $\Delta_k$ is an array of size 64$\times$64, where each element of the mask is set to -1 or 1 with 50\% probability.   Fig.~\ref{fig:SPI_vs_SelfGuiding_data} evaluates the error in the overlap of the true image and the guess image $(1-\langle \mathcal{O}|\sigma \rangle)$ as a function of iteration $k$.  We see that at every iteration the single-pixel and self-guiding approaches are equivalent, with the difference between the two less than~$\pm 10^{-12}$.

\clearpage
\newpage

\section{OSGQT} \label{S: SGQT Kaczmarz}
\vspace{-10pt}
\subsection{Experimental details} \label{S: exp details}
\vspace{-10pt}
\begin{figure*}
    \centering
    \includegraphics[width=1\linewidth]{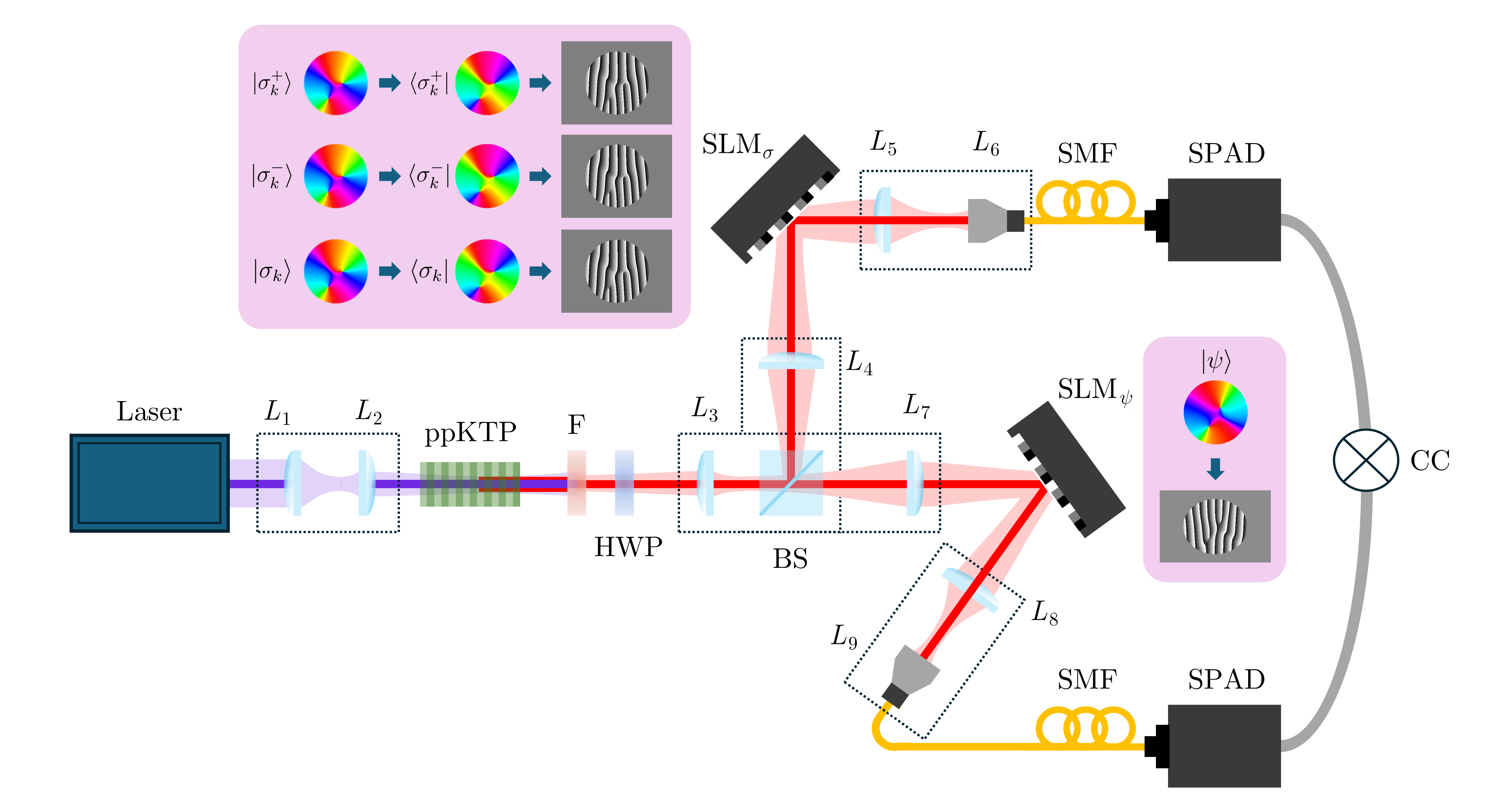}
    \caption{A 405 nm CW laser is focused into a ppKTP crystal to undergo type-I SPDC. The pump beam is blocked by a 550~nm long pass filter (F). The 810 nm entangled pairs pass through a half-wave plate (HWP) for polarisation optimisation before being separated via a beam splitter (BS). The signal and idler are imaged onto spatial light modulators (SLM) before each being coupled into single-mode fibre (SMF) connected to a single-photon avalanche diode (SPAD). SLM$_\psi$ imprints the true state $\ket{\psi}$ onto the signal phase and SLM$_\sigma$ the complex conjugate of $\ket{\sigma_k^+}, \, \ket{\sigma_k^-}, \ \mathrm{and}\ \ket{\sigma_k}$ onto the idler phase, corresponding to three different measurements for each iteration $k$. Note that although two separate SLMs are displayed, in the experiment, the photons were imaged onto two halves of one SLM. Coincidence counting (CC) is recorded using timetagging of the joint detection events. The dotted boxes represent 4$f$ imaging systems, with the lenses (L) having focal lengths $f_1$~=~200mm, $f_2 = f_3$~=~50~mm, $f_4$~=~$f_5$~=~$f_7$~=~$f_8$~=~500 mm, and $f_6$~=~$f_9$~=~2~mm.}
    \label{fig: exp}
\end{figure*}
Fig.~\ref{fig: exp} shows a schematic of the experimental set up. We image two entangled photons, generated by pumping a temperature-controlled type-I ppKTP crystal with a 405 nm CW laser, onto two halves of a Holoeye Pluto-2.1 SLM (displayed as individual SLM$_\psi$ and SLM$_\sigma$ in Fig.~\ref{fig: exp}). SLM$_\psi$ imprints the target $d=5$ OAM superposition $\ket{\psi}$ onto the phase of the signal. For each iteration~$k$, three measurements are performed corresponding to three different SLM$_\sigma$ projection holograms for the idler. The first two correspond to $|\sigma_k^\pm\rangle$ as part of SGQT algorithm, whereas the third one, corresponding to $\ket{\sigma_k}$, is merely to evaluate the performance of the SGQT algorithm. Note that SLM$_\sigma$ imprints the complex conjugate of phases as the signal and idler are anti-correlated in OAM.
The diffraction gratings, used such that only the first order can be selected, are confined to a circular mask that is slightly larger than the single-mode fibre collection mode. This is to limit an increase in coincidences that result from updating the estimate $\ket{\sigma_k}$ outside the spatial region corresponding to the true mode $\ket{\psi}$.
Joint projections are performed by coupling the photons in single-mode fibres connected to single-photon detectors. Coincidences are established using timetagging via a PicoQuant Hydraharp. 
Due to the photons being entangled in OAM, the recorded coincidence counts should be proportional to the fidelity $|\braket{\sigma|\psi}|^2$. 
We recorded with a mean coincidence rate $\approx$  5$\times 10^3 \ \mathrm{s}^{-1}$ 
when $\ket{\sigma} = \ket{\psi}$.
  
\subsection{Fidelity metrics for current best estimates $\sigma_k$} \label{S: Fidelity metrics}
\vspace{-10pt}
\begin{figure}[t]
    \centering
    \includegraphics{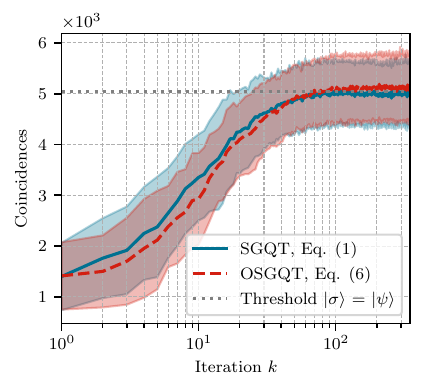}
    \caption{Comparison of recorded counts for SGQT (solid blue) and OSGQT (dashed red). The lines show the mean counts for 10 quantum states $\ket{\psi}$ recorded with a 1s exposure time, with the shaded regions representing the standard deviation. After a certain number of iterations, the detected coincidences for $\ket{\sigma_k}$ surpass the coincidence threshold that is measured when $\ket{\sigma}$ is set to $\ket{\psi}$.}
    \label{fig: counts vs k}
\end{figure}

\begin{figure}
    \centering
\includegraphics{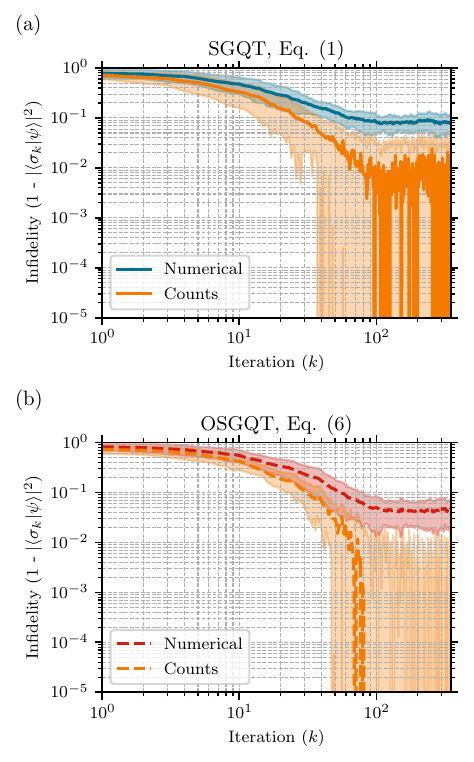}
    \caption{Comparison of the inferred infidelities of (a) SGQT and (b) OSGQT based on numerical phase overlap (solid blue or dashed red) or counts (solid or dashed orange). The lines show the mean infidelities for 10 quantum states $\ket{\psi}$ recorded with a 1s exposure time, with the shaded regions representing the standard deviation. After a certain $k$, we calculate nonphysical fidelities~($>100\%$) values when they are based on the detected counts.}
    \label{fig: Counts vs wavetrace infidelities}
\end{figure}

In a SGQT algorithm a current best estimate $\ket{\sigma_k}$ of true state $\ket{\psi}$ is updated every iteration after two measurements that evaluates the distance measure between $\ket{\psi}$ and $\ket{\sigma^\pm_k}$. At no point, $\ket{\sigma_k}$ is assessed against $\ket{\psi}$. However, in order to measure the performance of the algorithm we do evaluate $\ket{\sigma_k}$ against $\ket{\psi}$. Note that this in no way impacts the algorithm itself. Rambach \textit{et al.} \cite{rambach_robust_2021} estimate the fidelity between $\ket{\sigma_k}$ and $\ket{\psi}$ based the detected counts. However, this gives us a skewed metric with nonphysical values. After a certain $k$ our recorded coincidences surpass the counts we obtain for $\ket{\sigma} = \ket{\psi}$ as can ben seen in Fig.~\ref{fig: counts vs k}, which would represent nonphysical fidelities that are larger than 100$\%$. This is easily explained by the fact that SGQT updates $\ket{\sigma_k}$ in the direction that leads to more detection events regardless of $\ket{\psi}$. Walking $\ket{\sigma_k}$ towards $\ket{\psi}$ leads to more counts, but so can small corrections for alignment and aberrations. To evaluate the fidelity fairly, we numerically compute the overlap of the phase profiles. The phases are normalised using an intensity envelope corresponding to the phase area over which the algorithm evaluates the distance measure, i.e. detection events. Specifically, this envelope is given by the product of the circular mask defining the SLM modes and a phase-step scan \cite{srivastav_characterizing_2022} of the entangled modes, which accounts for both the mode size on the SLM and the collection mode size of the fibre. A comparison of the mean infidelity development based on counts and numerical phase overlap is shown in Fig.~\ref{fig: Counts vs wavetrace infidelities}. Note that the infidelity based on the numerical overlap is an underestimate because it does not account for misalignment, e.g. centre or size overlap of the SLM holograms with the modes, or prepare-and-measure imperfections. For both metrics, OSGQT outperforms SGQT.

\subsection{Scaling parameters $\alpha$ and~$\beta$} \label{S: scaling parameters}
\begin{figure*}
    \centering
    \includegraphics{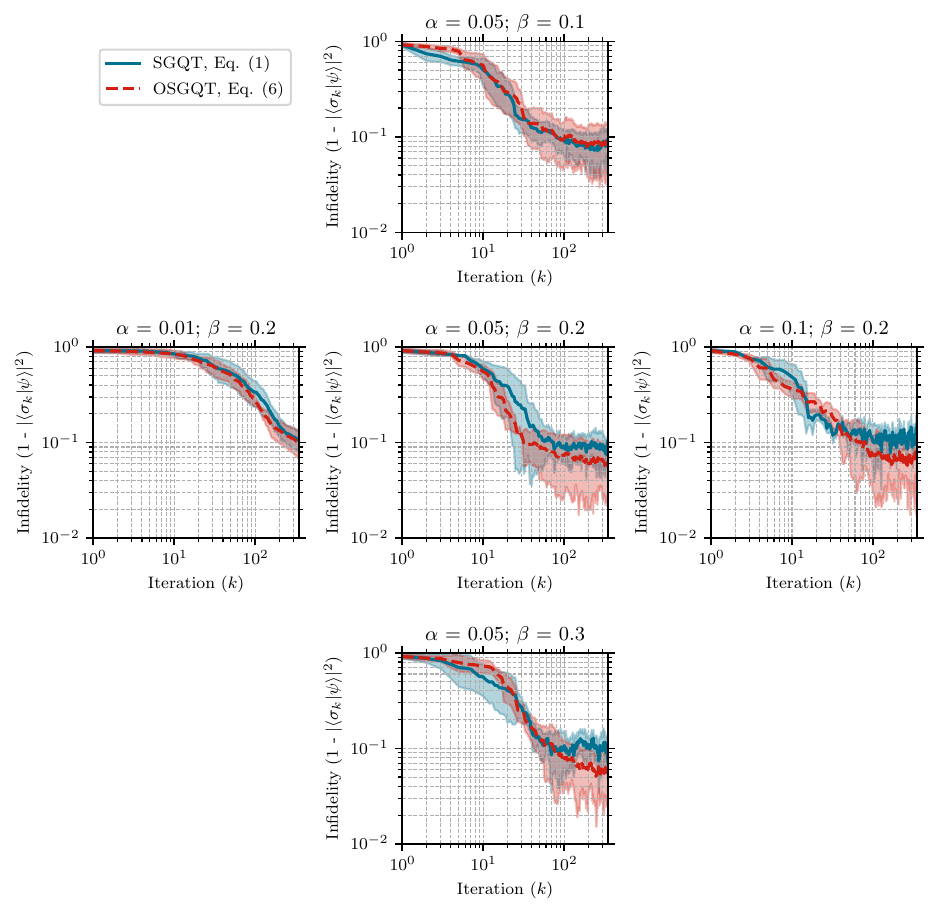}
    \caption{SGQT (solid blue) and OSGQT (dashed red) experimental performance for different scaling parameters $\alpha$ and $\beta$. The lines show the mean infidelities for three specific quantum states $\ket{\psi}$ with the shaded regions representing the standard deviation. Best SGQT (OSGQT) performance was found for $\alpha = 0.05;\ \beta=0.2.$ ($\alpha = 0.05;\ \beta=0.3$). Across all $\alpha$ and $\beta$ combinations, the best mean OSGQT result outperforms the best mean SGQT result.}
    \label{fig: alpha and beta combinations}
\end{figure*}

\vspace{-10pt}
In the original work on SGQT  \cite{RN1358}, scaling factors $\alpha$ and $\beta$ are defined to decay exponentially with the iteration index $k$ in order to achieve fast and accurate convergence. As the  estimate $\ket{\sigma_k}$ approaches the true state $\ket{\psi}$, progressively smaller updates to $\ket{\sigma_k}$ are required to prevent overshooting. As noted by Ferrie, the optimal choice of the scaling factors is highly problem-dependent and requires substantial resources to determine. Because we compare the fidelities at each $k$ of two algorithms, it seems a fairer comparison to use constant $\alpha$ and $\beta$, even though this limits the maximum fidelity that can be achieved. Even then, the two algorithms might benefit from different constant $\alpha$ and $\beta$. Therefore, we ran both algorithms for various values of scaling factors $\alpha$ and $\beta$, presented in Fig.~\ref{fig: alpha and beta combinations}. Comparing the best performance of each algorithm across different values of $\alpha$ and $\beta$ provides confidence that OSGQT indeed outperforms SGQT.

%

\subsection{Noise} \label{S: Noise}
\begin{figure*}
    \centering
    \includegraphics{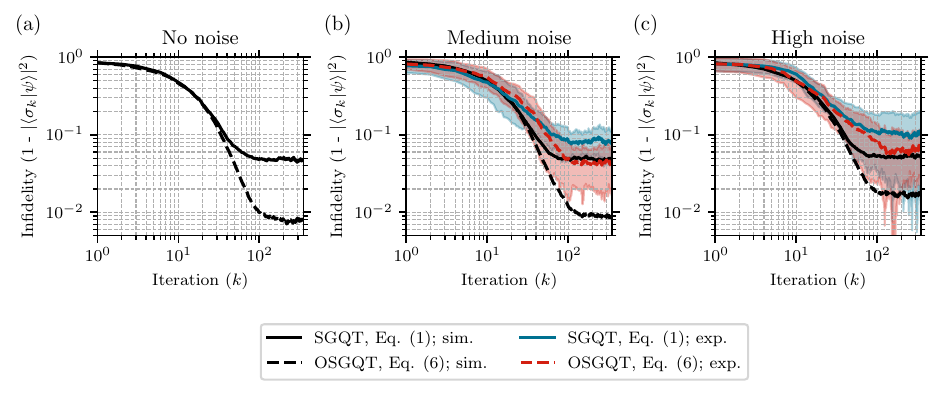}
    \caption{Numerical and experimental comparison of SGQT and OGQT performance for difference noise levels. Unknown quantum states $\ket{\psi}$ were $d=5$ OAM phase qudits, i.e. $\sum_{\ell=-2}^2c_\ell\ket{\ell}.$ Black (coloured) lines show the mean infidelity of 100 simulated (10 experimental) runs, with solid (dashed) lines representing SGQT (OSGQT). Shaded regions indicate the standard deviations for experimental results.
    (a) Numerical results only where no noise was added. (b) ((c)) Numerical and experimental results for an exposure time of 1s (0.1s) with a mean count rate of 5$\times10^3 \mathrm{s}^{-1}$ when $\ket{\sigma} = \ket{\psi}$. In simulation added noise is solely based on a Poisson distribution of the counts. Numerical and experimental results both demonstrate OSGQT reaches sub 0.1 infidelity thresholds faster on average and even allows to reach lower infidelities than the SGQT is able to regardless of number of iterations.}
    \label{fig: Noise_SGQT}
\end{figure*}

\vspace{-10pt}
Fig.~\ref{fig: Noise_SGQT} shows simulated and experimental performance of SGQT and OSGQT for different noise levels. 
In simulation, noise is added solely based on Poisson statistics of detection events, i.e. shot noise. 
The simulation computes the numerical fidelity between $\ket{\psi}$ and $\ket{\sigma_k^\pm}$ (similar as we evaluate the fidelity between $\ket{\psi}$ and $\ket{\sigma_k}$), labeled $F_\textrm{sim}$. 
The number of coincidences for $\ket{\sigma_\pm}$, labeled $N^\pm_\textrm{sim}$, is then simulated via
\begin{equation}
    N^\pm_\mathrm{sim} = P\left(F_\textrm{sim}^{\psi,\,\sigma^\pm}\times C_\textrm{exp}^{\sigma=\psi}\times I\right)
\end{equation}
where $P$ represents a Poisson distribution, $C_\textrm{exp}^{\sigma=\psi}$ is expected coincidence rate for when $\ket{\sigma} =\ket{\psi}$ (as measured prior in experiment), and $I$ is the integration time. As we were experimentally limited by the $C_\textrm{exp}^{\sigma=\psi}$ values we could achieve, different levels of shot noise were realised by varying the integration time.
In experiment, additional sources of noise include imperfections in state preparation, alignment, aberrations and detection.
Across the different noise levels, we still observe that OSGQT is able to achieve fidelities that are higher than those achievable by SGQT.

\newpage


%